\documentclass[a4paper,10pt]{article}

\pdfoutput=1

\usepackage{cite,amsmath,amsfonts,amsthm,fullpage}
\usepackage{youngtab}

\usepackage{graphics}
\usepackage{miniplot}
\usepackage{subfigure}

\newcommand{\Pa}{\mathop\mathrm{P}\nolimits}

\newcommand{\DP}{\mathop\mathrm{DP}\nolimits}
\newcommand{\OP}{\mathop\mathrm{OP}\nolimits}

\newcommand{\ttr}{\texttt{tr}}

\theoremstyle{plain}

\newtheorem{Lemma}{Lemma}
\newtheorem{Proposition}{Proposition}
\newtheorem{Corollary}{Corollary}
\newtheorem{Remark}{Remark}
\newtheorem{Example}{Example}

\theoremstyle{remark}

\def\Tr{\mathrm {Tr}}
\def\ttr{\mathrm {tr}}
\def\det{\mathrm {det}}

\def\diag{\mathrm {diag}}

\def\bp{\begin{Proposition}}
\def\ep{\end{Proposition}}
\def\bc{\begin{Corollary}}
\def\ec{\end{Corollary}}
\def\bl{\begin{Lemma}}
\def\el{\end{Lemma}}
\def\be{\begin{equation}}
\def\ee{\end{equation}}
\def\br{\begin{Remark}\rm\small}
\def\er{\end{Remark}}
\def\brs{\begin{remarks}.\\ \rm\
\begin{enumerate}}
\def\ers{\end{enumerate}\end{remarks}}
\def\bea{\begin{eqnarray}}
\def\eea{\end{eqnarray}}

\def\bx{\begin{Example}\rm\small}
\def\ex{\end{Example}}

\def\Tr{\mathrm {Tr}}
\def\tr{\mathrm {tr}}
\def\det{\mathrm {det}}

\def\diag{\mathrm {diag}}

\def\&{&{\hskip -20pt}}

\newcount\YDcount\YDcount=0
\def\YDsize{10pt}

\def\YD#1{%
\ifnum#1=0
 \ifnum\YDcount=0 \ifx\varnothing\undefined\emptyset\else\varnothing\fi
 \else\vskip1.4pt\egroup\YDcount=0\fi
\else
 \ifnum\YDcount=0 \YDcount=1\vcenter\bgroup\vskip1pt
 \else\nointerlineskip\fi
 \vbox{\hrule\hbox{\vrule height\YDsize
 \loop\hskip\YDsize\vrule\ifnum\YDcount<#1\advance\YDcount1\repeat}\hrule
 \kern-0.4pt}\expandafter\YD
\fi}

\usepackage[usenames,dvipsnames]{color}
\usepackage{ulem}

\def\pb{\mathbf{p}}
\def\tb{\mathbf{t}}

\def\xb{\mathbf{x}}
\def\yb{\mathbf{y}}

\begin{document}

\author{ E. N. Antonov \thanks{%
Petersburg Nuclear Physics Institute, Gatchina, RU-188350 St.Petersburg,
Russia, email: antonov@thd.pnpi.spb.ru},
A.Yu. Orlov\thanks{Institute of Oceanology, Nahimovskii Prospekt 36,
Moscow 117997, Russia;
NRC Kurchatov Institute, Moscow, 123182, Russia; 
Institute for Information Transmission Problems, 
Moscow 127051 Russia,
email: orlovs@ocean.ru}, 
D.V. Vasiliev\thanks{NRC Kurchatov Institute, Moscow, 123182, Russia; 
Institute for Information Transmission Problems, 
Moscow 127051 Russia, email: vasiliev@itep.ru
}}

\title{Coupling of different solvable ensembles of random matrices}

\date{January 30, 2022}
\maketitle

\begin{abstract}
Explicit expressions for multimatrix models with complex and unitary matrices allows
to couple these models with well-known unitary, orthogonsl and sympletic ensembles.
We consider examples of such mixed ensembles which are solvable in the sense that
the partition functions of such ensembles can be considered as tau functions of the classical
integrable equations.
\end{abstract}

\bigskip
\medskip


\section{Introduction \label{Introduction}}

If a matrix integral is a tau function as a function of its coupling constants we call it solvable. 
As far as we know the first solvable (in this sense) matrix model was presented in the preprint of \cite{GMMMO}
and other examples in \cite{KMMOZ} and \cite{GKM},\cite{KMMM},\cite{Kharchev}, and see also \cite{Mir_thesis}. Then we should point  out the work \cite{L1}. It is 
the direct continuation of \cite{O-2004-New} and also
of \cite{AOV} where the solvable cases were selected.

In works \cite{Belowezhie}, \cite{AOV}, \cite{Pogr} a number of solvable  models were
selected from the complex multi-matrix Ginibre ensemles (let us call such ensembles mCE) and from the models of many unitary
matrices (mUE) coupled to these multi-matrix Ginibre ensemles (mCUE). Here we explain that one can do one more step and couple the mentioned mCUE solvable models to the known ranodom matrix ensembles known as unitary, orthogonal and symplectic ones. Here we restrict ourself only with presentation the idea and giving an example:
we suggest the solvable model of the interaction of complex
matrices with the random Hermitian matrix which
was known to be of use in calculations in two-dimensional quantum gravity problems \cite{BrezinKazakov}.

\subsection{Complex and unitary multumatrix models}

The common feature of such models of solvable models 
is their perturbation series in coupling constants which look like as follows
\be\label{1}
Z_N(\pb,\pb^*)=\sum_\lambda r_\lambda(N) s_\lambda(\pb)s_\lambda(\pb^*)
\ee
where $N$ is the size of involved matrices. Apart of new models considered in
\cite{Belowezhie}, \cite{AOV},
similar perturbation series  was obtained in a number of different 
 models, among them: the 2-matrix model of Hermitian-antiHermitian matrices \cite{HO2003} 
(and as the result the very well-known one-matrix model introduced by Kazakov, Migdal, Brezin
and Gross), the model of two unitary matrices \cite{ZinZub}.

The series (\ref{1}) is known to be two-component KP tau function where
$\pb$ and $\pb^*$ play the role of the higher times \cite{KMMM},\cite{OS-TMP}.

In (\ref{1}) the sum ranges over the set of all partitions $\lambda=(\lambda_1,\dots,\lambda_N)$, $r_\lambda$ is
the so-called content product defined by the choice of the function (on $\mathbb{Z}$) $r$:
$$
r_\lambda(N)=\prod_{(i,j)\in\lambda} r(N+i-j)
$$
where the product ranges over all nodes of the Young diagram $\lambda$ with coordinates $i$ and $j$; the ``content'' of node is $j-i$.
The Schur polynomial $s_\lambda$ as the function of the variables $\pb=(p_1,p_2,\dots)$ (here, playing the role of coupling constants of the matrix model) is defined as follows:
$$
s_\lambda(\pb)=\det[s_{(\lambda_i-i+j)}(\pb)]_{i,j=1,\dots,N},\quad e^{\sum_{m>0}\frac 1m p_m z^m}=
\sum_{m\ge 0} z^m s_{(m)}(\pb)
$$
In case $\pb$ is given by 
$$
p_m=p_m(X)=\tr X^m
$$
we write $s_\lambda(X)$ instead of $s_\lambda\left(\pb(X)\right)$. In this case the definition
of the Schur polynomials can be re-written as
$$
s_\lambda(X)=\frac{\det \left[x_j^{\lambda_i-i+N} \right]_{i,j}}{\det \left[x_j^{-i+N} \right]_{i,j}}
$$
where $x_1,\dots,x_N$ are the eigenvalues of $X$,
which is the well-known formula for the character of $X\in \mathbb{GL}_N$.

Actually any embedded graph (or, the same, a fat graph, or a ribbon graph) drawn of the sphere and equipped with certain data gives rise
to the series (\ref{1}) which is known to be a tau function.

\paragraph{Embedded graph and the right hand side of (\ref{1})}

Embedded graph has only faces which homeomorphic to a disc. The boundaries of each face
are the sides of a ribbon edge. One assigns the positive (counter-clockwaise) orientation
to the boundaries of each face. It means that each ribbon edge consists of oppositely
directed arrows. One can say that each ribbon edge is a pair of glued oppositely directed arrows.
Thus, the boundary of the face is arrows successively placed next to each other with a positively chosen orientation.
Let us number each edge graph with positive numbers from 1 to $n$, where $n$ is the total number of edges. Then graph with $F$ faces possesses $V=2+n-F$ vertices. To the sides of each edge, say the edge number $i$ ($i>0$) we assign numbers $i$ and $-i$, thus, all sides of the graph are numbered
with numbers from the set $\pm 1,\dots, \pm n$. Let's denote this set $\mathfrak{N}$ and 
$\mathfrak{N}=2n$. Let us number the faces
of the graph with $1,\dots,F$.
By traversing each face, say $f_i$ in the positive direction, we obtain a set of numbers from the list indicated above, and such a set is defined up to a cyclic permutation; denote such set 
 by $\mathfrak{f}_i$ and call it the face cycle related to $f_i$. Next, we consider the symmetric group $S_{2n}$ which acts on $\mathfrak{N}$.
 Consider the element of $S_{2n}$ which is the product of the face cycles: 
 \be\label{face-cycle}
 (\mathfrak{f}_1)\cdots (\mathfrak{f}_F) \in S_{2n}
 \ee
 where the order of the factors is unimportant because by construction these are non-intersecting 
 cycles.
 
 Then, let us numerate the vertices of the graph with $1,\dots,V$.
 Next, consider the set of numbers around each vertex, say $v_i$, where we write down of 
 the numbers of
 the incoming arrows (thus, they belong to $\mathfrak{N}$) which we consequently write down traversing a vertex in the negative (or, the same, in the clockwise) direction. Such a set
defined up to the cyclic permutations we call the $i$-th vertex cycle and denote $\mathfrak{v}_i$.
The product of the (by definition) non-intersecting vertex cycles gives rise to the other element 
of $S_{2n}$:
$$
(\mathfrak{v}_1)\cdots (\mathfrak{v}_V) \in S_{2n}
$$
There is the known relation between these two elements related to any embedded graph:
\be\label{graph-combinatorial}
\sigma \circ (\mathfrak{f}_1)\cdots (\mathfrak{f}_F) =(\mathfrak{v}_1)\cdots (\mathfrak{v}_V)
\ee
where in the left hand side we have the composition of the involution $\sigma=\sigma^2$ without
fixed point and the product of face cycles. In our notations $\sigma$ acts as the sumulteneous
transposition $i \leftrightarrow -i$ for all $i=1,\dots,n$. Evidently it follows also
\be\label{graph-combinatorial*}
\sigma \circ (\mathfrak{v}_1)\cdots (\mathfrak{v}_V) =(\mathfrak{f}_1)\cdots (\mathfrak{f}_F)
\ee
This is a combinatorial description of the embedded graph\footnote{
In \cite{LZ} it written in a slightly different way in terms of the numbering of the half-edges of the dual graph.}, drawn on any surface with Euler characteristic $E=F-n+V$.
Formulas (\ref{graph-combinatorial}) and (\ref{graph-combinatorial*}) are related to dual graphs.

As it was noticed in \cite{NO2020tmp} we have a wondefful analogue of this relations in terms
of integrals over matrices.

\paragraph{Multi-matrix Ginibre ensemble, ensembles of unitary matrices, mixed ensembles.\label{mixed}}

The complex Ginibre ensemble is defined as the space of complex $N\times N$ matrices and by a measure on the space which is
\be
d\mu(Z)= e^{-N\Tr ZZ^\dag}\prod_{i,j\le N} d^2 Z_{i,j}
\ee
The $n_1$-matrix complex Ginibre ensemble is defined as the space of $N\times N$ matrices 
$Z_1,\dots,Z_{n_1}$ with the measure $d\mu(Z_1,\dots,Z_{n_1})=\prod_{i=1}^{n_1} d\mu(Z_i)$.

The ensemble of $n_2$ unitary matrices is defined by the space of 
$U_1,\dots,U_{n_2}\in \mathbb{U}_N$
with the measure $d\nu(U_1,\dots,U_{n_2})$ equal to the product of the Haar measures 
$d_*U_1\cdots d_*U_{n_2}$.

In what follows in this section we consider the mixed ensemble which contains $n=n_1+n_2$ matrices 
$Z_1,\dots,Z_{n_1}$ and $U_1,\dots,U_{n_2}$ with measure $d\Omega_{n_1,n_2}=d\mu(Z_1,\dots,Z_{n_1})d\nu(U_1,\dots,U_{n_2})$.
Let us denote the set of these $n_1+n_2$ matrices by $X$ and denote the measure by $dX$. The expectation of any function $f$ of the entries
of this set of matrices is defined as
$$
\langle f \rangle = \int_{\Omega_{n_1,n_2}} f
dX
$$
where $\Omega_{n_1,n_2}=\mathbb{GL}^{\otimes n_1}\otimes \mathbb{S}^{\otimes n_2}$.

If $\Gamma$ an embedded graph drawn on the sphere $\mathbb{S}^2$ which has $n$ ribbon edges
and $2n$ sides of these edges numbered by $\mathfrak{N}$ as it was described in the previous paragraph. Then we consider the equipped graph: to a side of the edge numbered with $i$ (we recall that $i\in\mathfrak{N}$) we match the matrix $X_i$ while to the opposite side of the same edge
which is numbered with $-i$ we match the Hermitian conjugate matrix $X_{-i}:=X_i$.
Here each $X_i$ is one of the matrices $Z_1,\dots,Z_{n_1},U_1,\dots,U_{n_2}$. Thus, the matrices 
$X_1,\dots,X_n$ belong to our mixed ensemble.

Apart of these $2n$ matrices we introduce different $2n$ matrices $C_{\pm 1},\dots,C_{\pm n}$ which contribute the equipping of $\Gamma$ as follows. Let us give numbers to the corners by assigning
to it the number of the side which income to this corner.

Consider a cycle $\mathfrak{f}_i=(j_1,j_2,\dots,j_{\ell_i})$ where $\ell_i$ is the number of edges
of the face $f_i$.
To each face cycle $(\mathfrak{f}_i)$ we relate the cycle product
$$
\mathfrak{f}_i\,\to\,{\cal F}_i
$$
which is the product $C_{j_1}C_{j_2}\cdots C_{j_{\ell_i}}$. Let us also introduce 
the product of the pairs of matrices according to
\be
{\cal F}_i(X)= X_{j_1}C_{j_1}\cdots X_{j_{\ell_i}}C_{j_{\ell_i}}
\ee
We call the replacements $C_j\to X_jC_j=:C_j(X)$, where each of source matrices is 
multiplied by the random matrix with the same number from the left dressing procedure.

We call such product the face monodromy because it is the product of matrices 
obtained as the product of the matrices that we encounter when traversing the 
edge boundary in the positive direction.

In a similar way, we obtain the monodromy of vertices - this is the product of matrices corresponding to the cycle of vertices:
$$
\mathfrak{v}_i\,\to\,{\cal V}_i
$$
Say, the cyclic product related to the vertex cycle
$\mathfrak{V}_i = (k_1,\dots,k_{\ell(i)})$ where $\ell_i$ is the valency of the vertex $i$
is  $C_{k_1}\cdots C_{k_{\ell_i}}$.

We have a wonderful relation \cite{NO2020tmp} 
\be\label{I}
\int_{\Omega_{n_1,n_2}} \prod_{a=1}^F s_{\lambda^a}\left({\cal F}_a(X)\right) dX=\delta_\lambda
N^{-n_1d}\left(s_\lambda(\pb_\infty)\right)^{-n_1}\left(s_\lambda(\mathbb{I}_N)\right)^{-n_2}
\prod_{a=1}^V s_{\lambda^a}\left({\cal V}_a\right)
\ee
and 
\be\label{II}
\int_{\Omega_{n_1,n_2}} \prod_{a=1}^V s_{\lambda^a}\left({\cal V}_a(X)\right) dX=
\delta_\lambda  N^{-n_1 d}
\left(s_\lambda(\pb_\infty)\right)^{-n_1}\left(s_\lambda(\mathbb{I}_N)\right)^{-n_2}
\prod_{a=1}^F s_{\lambda^a}\left({\cal F}_a\right)
\ee
where $\delta_\lambda$ is equal to 1 in case $\lambda^1=\cdots = \lambda^F$ and where we
denote $\lambda^1$ by $\lambda$. And $\delta_\lambda=0$ otherwise. 
Here
\be
s_\lambda(\pb_\infty)=\frac{d_\lambda}{d!},\quad d_\lambda=
\frac{\prod_{i<j\le N}(\lambda_i-\lambda_j-i+j)}{\prod_{i=1}^N(\lambda_i-i+N)!},\quad 
d=|\lambda|
\ee
(where $d_\lambda$ is the dimension of the representation $\lambda$ of the symmetric group $S_d$)
and
\be
s_\lambda(\mathbb{I}_N))=\frac{s_\lambda(\pb_\infty)}{(N)_\lambda},\quad
(N)_\lambda=\sum_{(i,j)\in\lambda}(N-i+j)=\prod_i \frac{\Gamma(N+\lambda_i-i)}{\Gamma(N-i)}
\ee
which is the dimension of the representation $\lambda$ of the linear group $\mathbb{GL}_N$.
Formulas (\ref{I}) and (\ref{II})
can be related respectively to (\ref{graph-combinatorial}) and 
(\ref{graph-combinatorial*}).

\paragraph{Matrix models.}
Now we consider the matrix model related to an embedded graph $\Gamma$ as
$$
Z(\tb,{C})=\int_{\Omega_{n_1,n_2}} \prod_{i=1}^F e^{\sum_{m>0}\frac 1m p^{(i)}_m \tr \left({\cal F}_i(X)\right)^m}dX=
$$
\be\label{ZU-MM}
\sum_{\lambda\atop \ell(\lambda)\le N} N^{-n_1|\lambda|} \left((N)_\lambda\right)^{-n_2} 
\left(s_\lambda(\pb_\infty)\right)^{-n}
\prod_{i=1}^F s_\lambda(\pb^i)
\prod_{i=1}^V s_\lambda({\cal V}_i)
\ee
where $\tb$ denotes the collection of coupling constants $\pb^i,\,i=1,\dots,F$ and
${\cal C}$ denotes the collection of the source matrices $C_{\pm i},\,i=1,\dots,n$.
This model has $F$ sets of coupling constants $\pb^i=(p^{(i)}_1,p^{(i)}_2,\dots)$ and 
the set of $V$ independent combinations ${\cal V}_i$ constructed from the source matrices
(or, the same, from the corner matrices of the graph $\Gamma$.)

\paragraph{Specifications to the integrable family}
We obtain an integrable model in case the right hand side of (\ref{ZU-MM}) coinsides with
the right hand side of (\ref{1}).
In this case all Schur polynomials except a chosen pair (as in (\ref{1})) should contribute
to a content product as we find as the prefactor in (\ref{1}).
This is possible if we put some of $\pb^i$ equal to $\pb_\infty$ , or if we use the relations
\be\label{p(a)}
s_\lambda(\pb(a))=s_\lambda(\pb_\infty)\prod_{(i,j)\in\lambda}(a+j-i),\quad 
\pb(a)=(a,a,\dots)
\ee
or/and 
\be\label{poorX}
s_\lambda(X)=s_\lambda(\pb_\infty)\prod_{(i,j)\in\lambda}(N-m+j-i)
\ee
The last one is obtained in case the spectrum of $X\in\mathbb{GL}_N$ consists of $m$ unities 
and $N-m$ zeros. Let us note that in this case $s_\lambda(X)$ is equal to zero in if
the length of $\lambda$ exceeds $N-m$.

Let us note that the usage of (\ref{poorX}) means that in fact we deal with the rectangular metrices. The specification of $\pb^i$ means that
\be\label{p-infty}
{\rm for}\,\, \pb^i=\pb_\infty\,\, {\rm we \,have}\,\,
e^{\sum_{m>0}\frac 1m p^{(i)}_m \tr \left({\cal F}_i(X)\right)^m}=
e^{ \tr {\cal F}_i(X)}
\ee
\be\label{p(a)}
\qquad\quad{\rm for}\,\, \pb^i=\pb(a_i)\,\, {\rm we \,have}\,\,
e^{\sum_{m>0}\frac 1m p^{(i)}_m \tr \left({\cal F}_i(X)\right)^m}=
\det\left(1-{\cal F}_i \right)^{-a_i}
\ee
where $a_i$ is a chosen parameter.

Let us produce such specifications and obtain the $ Z(\tb,{\cal C},n)$ as the tau function which depends either on $\pb$ and ${\cal V}$:
\be\label{ZtV}
{\cal Z}_N(\pb,{\cal V})=\sum_{\lambda} r_\lambda(N)s_\lambda(\pb) s_\lambda({\cal V})
\ee
or on a pair of corner matrices, say, ${\cal V}_1$ and ${\cal V}_2$:
\be\label{ZVV}
{\cal Z}_N({\cal V}_1,{\cal V}_2)=\sum_{\lambda} r_\lambda(N)s_\lambda({\cal V}_1) s_\lambda({\cal V}_2)
\ee
with certain $r$ obtained after a chosen specification. Here ${\cal V}$ and ${\cal V}_{1,2}$ 
are any ones chosen from the set of vertex monodromies $\{{\cal V}_i$ where $i=1,\dots,V  \}$
(we remind that the rest part should be chosen to be equal matrices whose spectrum consists
of $1$ and $0$);
and for $\pb$ we choose one in the set of coupling constants $\{\pb^i,\,i=1,\dots,F\}$,
say $\pb=\pb^{(1)}$
(while the rest coupling constants are chosen according to (\ref{p-infty}) or (\ref{p(a)})).

\section{Coupling of the integrable mCUE with well-known solvable matrix models}

In series (\ref{ZtV}),(\ref{ZVV}) we are going to treat ${\cal V},{\cal V}_{1,2}$ as
random matrices which belong to enesembles different of the considered CUE ones.
The main example is the coupling with integrable ensembles of Hermitian matrices
known as one- and two-matrix models.

\paragraph{Modified one-matrix models with $d\mu(H)$}

The most known solvable matrix models are the models related to three classical
Wigner-Dyson ensembles: unitary, orthogonal and symplectic ones. Each of these matrix models
can be treated as tau functions where the coupling constants of these models turn out 
to be the higher times of these functions \cite{JM}. It is a special case of the KP tau function for the unitary ensemble, see \cite{GMMMO}. 
And it is an example of the tau function of the large BKP hierarchy\footnote{The large, or charged
BKP tau function was introduced in \cite{KvdLbispec}. The DKP tau function was introduced in 
\cite{JM}. Actially the DKP hierarchy of integrable equations can be incorporated into the large BKP hierarchy.} in case of the orthogonal ensemble \cite{L1} and an example of the DKP tau function in case of the symplectic ensemble \cite{L1}.

Matrix model related to the unitary Wigner-Dyson ensemble is as follows
\be
I_1(\pb)=\int e^{N\sum_{m>0} \frac 1m \tr(H^m)} d\nu(H) = 
\sum_{\lambda} \langle s_\lambda(H)\rangle s_\lambda(\pb)
\ee
where $H$ is a Hermitiam matrix
\be
\langle s_\lambda(H)\rangle =
\int s_\lambda(H) d\nu(H)
\ee
where $H$ is a Hermitiam matrix. The measure on the space of Hermitian matrices is
the Gaussian one:
\be\label{1MMmeasure}
d\nu(H)= e^{-\frac 12 N\tr (H)^2}dH,\quad dH=\prod_{i\le j} d\Re M_{i,j}\prod_{i < j} d\Im M_{i,j}
\ee
As we mentioned it is well-known that $I_1(\pb)$ is the KP tau function with respectivelyto the coupling constants $\pb=(p_1,p_2,\dots)$. As it is also well-know, any KP tau function
can be written is Sato form \cite{Sato} as
\be\label{Sato}
\tau^{\rm KP}(\pb)=\sum_\lambda \pi_\lambda s_\lambda(\pb)
\ee
where the numbers $\pi_\lambda$ solves certain bilinear relations called Plucker relations.
These numbers are known as the Plucker coordinates. For instance the numbers
$r_\lambda(N)s_\lambda(\pb^*)$ solve Plucker relation for any choice of the function $r$ 
and for any choice of $\pb^*$.

The fact which is of importance is that
if the numbers $\pi_\lambda$ solve Plucker relations then the numbers
$$
\tilde{\pi}_\lambda = \pi_\lambda r_\lambda(N)
$$
also solves Plucker relations for any choice of $r$. The easiast way to prove it is to use 
free fermions, see Appendix. It results to the fact that one can replace $\pi_\lambda$
by $\tilde{\pi}_\lambda$ in (\ref{Sato}) to get new KP tau function $\tilde{\tau}$.

Therefore, since $\langle s_\lambda(H) \rangle$ is the example of the Plucker coordinate,
we can apply it to the series
\be\label{Z-solvable}
Z_N(\pb,H)=\sum_\lambda r_\lambda(N) s_\lambda(\pb)s_\lambda(H)
\ee
and obtain the result that
\be
\langle Z_N(\pb,H) \rangle_{\cal H} = {\cal Z}_N(\pb)
\ee
is the example of the KP tau function.
We call it the coupling of the ensemble mCUE and the unitary ensemble.
And as we see this new ensemble is also solvable.

The similar coupling of mCUE can be done if we take two other different Wigner-Dyson
ensembles.

The large BKP tau function can be written in form
\be\label{BKP}
\tau^{\rm BKP}(\pb)=\sum_\lambda K_\lambda s_\lambda(\pb)
\ee
where $K_\lambda$ are called the Cartan coordinates, see \cite{KvdLbispec} and 
\cite{HB} which are defined as certain Pfaffians.
Examples of such series one can found in \cite{OST-I}, \cite{OST-II}.

Again, we have the property that if $K_\lambda$ are the Cartan coordinated then
\be
\tilde{K}_\lambda = K_\lambda r_\lambda(N)
\ee
also are the Cartan coordinates and the related $\tilde{\tau}^{\rm BKP}$ is the large 2BKPtau function. Thus, the coupling of a solvable ensemble mCUE with orthogonal and with symplectic
Wigner-Dyson ensembles:
\be
\langle Z_N(\pb,O) \rangle_O,\quad \langle Z_N(\pb,S) \rangle)S
\ee
yields BKP tau functions where $\pb$ is the set of the large BKP higher times.

In the same way we can couple any solvable ensemble mCUE with real and quaternionic
Ginibre ensembles because they are examples of the large BKP tau functions, see \cite{OrlovGinibre}
and with interpolating ensemble, see \cite{OST-II}.

\paragraph{Examples}.
The simplest example is the following model
\be
I_1=\int e^{\sum_{m>0}\frac 1m p_m\tr (HZI_nZ^\dag)}d\nu(H)d\mu(Z)
\ee
This is the simplest one-face graph related to the interval with vertices,
where one corner matrix is the Hermitian matrix $H$ and the other corner
matrix is $I_n=\diag(1,1,\dots,1,0,\dots,0)$.
We get
\be
I_1= \sum_\lambda  (N-n)_\lambda s_\lambda(\pb)  \int s_\lambda(H)   d\nu(H)
\ee

\paragraph{Modification of two-matrix models. Measure $d\rho(H_1,H_2)$.} 
Two-matrix models
A typical perturbation series for a solvable two-matrix models looks as
\be\label{2MM-Takasaki}
{\cal Z}_{{\cal V}\times{\cal V}'}=\sum_{\lambda,\mu} s_\lambda(\tb) g_{\lambda,\mu}(N) s_\mu(\tb)
\ee
where $ g_{\lambda,\mu}$ is the mean of the pair of the Schur functions in a certain 
ensemble of two random matrices:
\be
 g_{\lambda,\mu}(N)=\langle s_\lambda(H_1)s_\mu(H_2) \rangle_{H_1,H_2}=
 \int s_\lambda(H_1)s_\mu(H_2) d\rho(H_1,H_2)
\ee

Let us take
\be
d\rho(H_1,H_2)=e^{-a\tr (H_1)^2-a\tr(H_2)^2 +c\,\tr (H_1H_2)}dH_1dH_2
\ee
where $dH$ were defined in (\ref{1MMmeasure}),
as the main example.

\br The examples were written down in
\cite{HO2003},\cite{OrlovShiota} as 
$$
d\rho(H_1,H_2)=\tau_r(H_1H_2,\mathbb{I}_N)d\mu(H_1)d\mu(H_2)
$$
where $\tau_r$ is a KP tau function of the hypergeometric type.
\er

For instance, consider the standard model of two interacting Hermitian matrices:
\be\label{2MM}
J_N(\tb^1,\tb^2)= \int e^{V(H_1,\tb^{1})}e^{V(H_2,\tb^{2})}d\rho(H_1,H_2)
\ee
\be
V(H_i,\tb^{i})=\sum_{m>0} \frac 1m p^{(i)}_m \tr\left( H_i \right)^m.
\ee
Here, the prefactor $g_{\lambda,\mu}$ is the determinant of the matrix of moments:
\be\label{two-matrix-kern}
g_{\lambda,\mu}(N) =\det \left[\int\int x^{h_i}e^{-ax^2-ay^2 +cxy}y^{h'_j}\right]_{i,j},\quad h_i=\lambda_i-i+N,\quad h'_j=\mu_j-j+N
\ee
with certain reasonable parameters $a$ and $c$ to get finite $g_{\lambda,\mu}$.
Such determinantal form of the prefactor provides the series in the left hand side of 
(\ref{2MM-Takasaki}) to be a tau function, where $\tb^1$ and $\tb^2$ each play the role
of the set of the KP higher times,   see \cite{Takasaki}.

In case we make the replacement 
\be
V(H_i,\tb^{i})\,\to\,{\cal Z}^{(i)}(H_i,\tb^{(i)}),\quad i=1,2
\ee
where ${\cal Z}^{(i)}(H_i,\tb^{(i)})$ are two different matrix integrals (\ref{ZtV}):
\be
{\cal Z}^{(i)}_N(H_i,\tb^{(i)}) = 
\sum_{\lambda \atop \ell(\lambda)\le N} r^{(i)}_\lambda s_\lambda(H_i)s_\lambda(\tb^{(i)})
\ee
we obtain the coupled model whose partubtaion series has the form (\ref{2MM-Takasaki})
where one should make the replacement
\be\label{g-tilde}
g_{\lambda,\mu}\,\to\,\tilde{g}_{\lambda,\mu}(N)r^{(1)}_\lambda(N)r^{(2)}_\mu(N)
\ee
\be\label{ZZ}
\langle {\cal Z}_N^{1}(H_1,\tb^1){\cal Z}_N^{2}(H_2,\tb^2)\rangle_{H_1,H_2}=\sum_{\lambda\atop \ell(\lambda)\le N}
s_\lambda(\tb^1)\tilde{g}_{\lambda,\mu} s_\mu(\tb^2)
\ee
It is easy to verify that for any choice of functions $r^{(1)}$ and $r^{(2)}$,
the prefactors (\ref{g-tilde}) in the right hand side of (\ref{ZZ}) has the form of the determinant of a matrix,  which is equal to the moment matrix multiplied by diagonal ones from the left and 
from the right. Then it follows that the right hand side of (\ref{ZZ}) is also a tau function.

Perhaps, the simplest example (\ref{ZZ})  is
\be
J_N(\tb)=
\int 
e^{\sqrt{-1}H_1H_2+\sum_{i=1,2}\sum_{m>0}\frac 1m t^{(i)}_m\tr\left(H_i Z^\dag I_{n_i}Z \right)^m }
dH_1 dH_2 d^2 Z
\ee
\be\label{Nnn}
=\sum_{\lambda\atop\ell(\lambda)\le N} (N)_\lambda(N-n_1)_\lambda(N-n_2)_\lambda s_\lambda(\tb^1)s_\lambda(\tb^2)
\ee
(we take $a=0,\,c=\sqrt{-1}$ in (\ref{two-matrix-kern})).
The perturbation series (\ref{Nnn}) is the example of (the so-called hypergeomteric) KP tau functions with respect 
to any set: either $\tb^{1}$, or $\tb^{2}$.
Say, if one choose $\tb^1=(1,1,1\dots)$ and $\tb^2=(x,x^2,x^3,\dots)$ he obtains the divergent hypergeometric
function ${_3}F_0(N,N-n_1,N-n_2;x)$.
If only a finite set among $\{t^{(i)}_m,\,m=1,2,\dots\}$ is not zero valued, then the series in the right hans side diverges.
One can show the series in the left hans side converges in an open set of parameters $\tb^i$.
One can notice that if we take $p^{(1)}_m=-\sum_{i=1}^M y_i^m,\, m=1,2,\dots$ the series (\ref{Nnn}) is a polynomial of at most degree $NM$ if we put $\deg p^{(1)}_m=m$.

\paragraph{Chains of integrals.} If one take the following set of integrals (\ref{ZtV}),(\ref{ZVV})
$$
{\cal {Z}}(\pb^{(1)},{\cal {V}}_1),\, {\cal Z}({\cal V}_2,{\cal V}_3)  ,\dots,\,{\cal Z}({\cal V}_n,\pb^{(2)})
$$
he can tie them in  a chain matrix model with the help of set of measures 
$d\rho({\cal V}_1,{\cal V}_2),\,d\rho({\cal V}_1,{\cal V}_2),\dots$:
$$
\int {\cal {Z}}(\pb^{(1)},{\cal {V}}_1),\, {\cal Z}({\cal V}_2,{\cal V}_3)  ,\dots,\,{\cal Z}({\cal V}_n,\pb^{(2)}) d\rho({\cal V}_1,{\cal V}_2)d\rho({\cal V}_3,{\cal V}_4)\cdots 
d\rho({\cal V}_{2n-1},{\cal V}_{2n}))
$$
This is also a solvable matrix model with the two sets of higher times $\pb^{(1,2)}$.
Examples will be considered in the more detailed text.

\bigskip
\bigskip
\noindent
\small{ {\it Acknowledgements.}
 The work of A.Orlov was supported by the Russian Science
Foundation (Grant No.20-12-00195).

\bigskip



\appendix

\section{Partitions. The Schur polynomials}
 
We recall that a nonincreasing set of nonnegative integers $\lambda_1\ge\cdots \ge \lambda_{k}\ge 0$,
we call partition $\lambda=(\lambda_1,\dots,\lambda_{l})$, and $\lambda_i$ are called parts of $\lambda$.
The sum of parts is called the weight $|\lambda|$ of $\lambda$. The number of nonzero parts of $\lambda$
is called the length of $\lambda$, it will be denoted $\ell(\lambda)$. See \cite{Mac} for details.
Partitions will be denoted by Greek letters: $\lambda,\mu,\dots$. The set of all partitions is denoted by
$\Pa$. The set of all partitions with odd parts is denoted $\OP$.
Partitions with distinct parts are called strict partitions, we prefer
letters $\alpha,\beta$ to denote them. The set of all strict partitions will be denoted by $\DP$.
The Frobenius coordinated $\alpha,\beta$ for partitions $(\alpha|\beta)=\lambda\in\Pa$ are of usenames
(let me recall that the coordinates $\alpha=(\alpha_1,\dots,\alpha_k)\in\DP$ consists of the lengths of arms counted
from the main diagonal of the Young diagram of $\lambda$ while
$\beta=(\beta_1,\dots,\beta_k)\in\DP$ consists of the lengths of legs counted
from the main diagonal of the Young diagram of $\lambda$, $k$ is the length of the main diagonal of $\lambda$,
see \cite{Mac} for details).

To define the Schur function $s_\lambda$, $\lambda\in\Pa$ at the first step we introduce the 
set of elementary Schur functions $s_{(m)}$ by
$$
e^{\sum_{m>0}\frac 1m p_m x^m}=\sum_{m\ge 0} x^m s_{(m)}(\pb)
$$
where the variables $\pb=(p_1,p_2,p_3,\dots)$ are called power sum variables. For 
$\lambda=(\lambda_1,\lambda_2,\dots)\in\Pa$
we define 
\be
s_\lambda(\pb)=\det\left[s_{(\lambda_i-i+j)}(\pb)  \right]_{i,j>0}
\ee
If we put $p_m=p_m(X)=\ttr X^m$ where $X$ is a matrix we write $s_\lambda(\pb(X))=s_\lambda(X)$.

 \paragraph{Examples of tau functions}

 \paragraph{Vacuum tau functions}

Notice that
$$
e^{\sum_m t_m\ttr X^m}\quad {\rm and}\quad e^{-\ttr XY}
$$
can be viewed as simplest (``vacuum'') Toda lattice tau function !

\paragraph{Sato-Takasaki series}

2KP (-Toda lattice) tau function,
(Sato,Takasaki,Takebe):
$$
\tau_N({\bf t},{\bf t}')=\sum_{\lambda,\mu} s_\lambda({\bf t}) g_{\lambda,\mu}(N) s_\mu({\bf t}')
$$
where 

 \begin{itemize}
  \item sum ranges over all possible pairs of partitions $\lambda,\mu$
  \item $s_\lambda , s_\mu$ are the Schur function (polynomials in KP higher times)
  \item $g_{\lambda,\mu}$ is the determinant of a certain matrix (initial data for the TL solution)  
 \end{itemize}
 
{\bf Example}
$$
e^{\sum_m t_m \ttr X^m}=\sum_{\lambda} s_\lambda({\bf t})s_\lambda(X)
$$
$$
e^{\ttr XY}=\sum_{\lambda} s_\lambda(XY) s_\lambda(1,0,0,\dots)
$$

\section{Matrix ensembles \label{Ensembles}}

\paragraph{Unitary matrices.}
The Haar measure on $\mathbb{U}_N$ in the explicit form is written as
\be
d_*U=\frac{1}{(2\pi)^N}\prod_{ 1\le i<k\le N} |e^{\theta_i}-e^{\theta_k}|^2 \prod_{i=1}^N d\theta_i,\quad 
-\pi < \theta_1 < \cdots <\theta_N \le \pi
\ee

\paragraph{Hermitian matrices.}

This case can be considered in the same way as the prevous one. 
It is natural \cite{Mehta} to take the following measure on the space $\mathfrak{H}_N$ of Hermitian $N\times N$ matrices:
$$
 d\Omega_{N,w}(X)=C_N\prod_{i\le j\le N}e^{-w\left(\Re X_{i,j}\right)^2 }d\Re X_{i,j} 
 \prod_{i < j\le N} e^{-w\left(\Im X_{i,j}\right)^2 } d\Im X_{i,j}  
$$
$$
 = c_N \left(\Delta(\xb)\right)^2  \prod_{i=1}^N e^{-w x^2_i}dx_i d_*U
$$
where $w>0$ is a parameter and where we use 
 $X=U\diag\left(\,\,\,0\quad x_i\atop -x_i\,\,0\right)U^{-1}$, $U\in \mathbb{U}_N$.

\br
In many applied problems in which random matrices are used, it is very convenient to assume 
that the parameter $w$ is proportional to the size of the matrices $N$.
\er
  
 \paragraph{Orthogonal matrices.}
The Haar measure on $\mathbb{O}_N$ is 
\be
d_*O=\begin{cases}
\frac{2^{(n-1)^2}}{\pi^n}\prod_{i<j}^n \left(\cos(\theta_i)-\cos(\theta_j)^2  \right)\prod_{i=1}^{n},\quad N=2n
      \\
\frac{2^{n^2}}{\pi^n}\prod_{i<j}^n \left(\cos(\theta_i)-\cos(\theta_j)^2  \right)\prod_{i=1}^{n}
 \sin^2\frac{\theta_i}{2}  d\theta_i ,\quad N=2n+1
     \end{cases}
\ee 
 where
$d\theta_i ,\quad 0\le \theta_i <\cdots<\theta_n \le \pi$. The prefactors are chosen to provide 
$\int_{\mathbb{O}_N} d_*O=1$.
 
By $\mathfrak{S}_N$ we denote the space of $N\times N$ real skew-symmetric matrices.
Recall that the eigenvalues of real skew fields are purely imaginary;
the eigenvalues occur in pairs $\pm x_i\sqrt{-1},\,i=1,\dots,n$  in the case $N=2n$ while
in the case $N=2n+1$ there is an additional eigenvalue $x_{2n+1}=0$.

\paragraph{Skew-symmetric matrices.}
For any $X\in\mathfrak{S}_N$ there exists such $O\in\mathbb{O}_N$ that
$X=O\diag\left\{\left(\,\,\,0\quad x_i\atop -x_i\,\,0\right)_{i=1,\dots,n}\right\}O^{-1}\in\mathfrak{S}_{2n}$,
where $O\in \mathbb{O}_{2n}$,
$X=O\diag\left\{\left(\,\,\,0\quad x_i\atop -x_i\,\,0\right)_{i=1,\dots,n},0\right\}O^{-1}\in\mathfrak{S}_{2n+1}$,
where $O\in \mathbb{O}_{2n+1}$.

The measure on $\mathfrak{S}_N$ is as follows:
$$
d\omega_N(X)=\pi^{\frac12 (n-n^2)}\prod_{i<j}^n e^{-X_{i,j}^2}dX_{i,j}
$$
\be\label{skew_symm_measure}
=\prod_{i<j\le n  } (x^2_i - x^2_j)^2  \prod_{i=1}^n e^{-x^2_i}dx_i d_*O
\begin{cases}
c_1,\quad N=2n
\\
c_2 x^2_i,\quad N=2n+1
\end{cases}
\ee
where prefactors $c_1=2^{-n}\prod_{i=0}^{n-1} \left( \Gamma(2+i)\Gamma(\tfrac12+i) \right)^{-1}$ and 
$c_2=2^{-n}\prod_{i=0}^{n-1} \left( \Gamma(2+i)\Gamma(\tfrac32+i) \right)^{-1}$ 
provide the normalization $\int d\omega_N(X)=1$, see (17.6.5) in \cite{Mehta}.

It is known \cite{BrezinHikami} that 
\be\label{IZHC_O}
\int_{\mathbb{O}_N} e^{\ttr \left(OXO^{-1}Y\right)} d_*O =\begin{cases}
c_1 \frac{\det\left[2\cosh(2x_iy_i) \right]_{i,j}}{\Delta(\xb^2)\Delta(\yb^2)},\quad N=2n
                                              \\  
c_2 \frac{\det\left[2\sinh(2x_iy_i) \right]_{i,j}}{\Delta(\xb^2)\Delta(\yb^2)\prod_{i=1}^n x_iy_i },\quad N=2n+1                                          
                                             \end{cases}
\ee
where $X,Y\in\mathfrak{S}_N$ and $\pm x_i\sqrt{-1}$ and $\pm y_i\sqrt{-1}$ are eigenvalues of respectively 
$X$ and $Y$.


\paragraph{Complex matrices.}
The measure on the space of complex matrices is defined as 
\be
d\Omega(Z)=c_N \prod_{a,b=1}^N d\Re Z_{ab}d\Im Z_{ab}\text{e}^{-N|Z_{ab}|^2}
\ee

We will consider integrals over $N\times N$ complex matrices $Z_1,\dots,Z_n$ where the measure is defined as
\be\label{CGEns-measure}
d\Omega(Z_1,\dots,Z_n)= c_N^n
\prod_{i=1}^n\prod_{a,b=1}^N d\Re (Z_i)_{ab}d\Im (Z_i)_{ab}\text{e}^{-N|(Z_i)_{ab}|^2}
\ee
where the integration domain is $\mathbb{C}^{N^2}\times \cdots \times\mathbb{C}^{N^2}$ and where $c_N^n$
is the normalization
constant defined via $\int d \Omega(Z_1,\dots,Z_n)=1$.

\end{document}